\documentclass[aps,pra,twocolumn,noshowpacs]{revtex4}
\usepackage{graphicx,epsfig}
\usepackage{amsbsy}
\usepackage{amsmath}
\graphicspath{{figures//}}

\begin{document}
\title{Spin-Nematic Squeezed Vacuum in a Quantum Gas}
\author{C.D. Hamley, C.S. Gerving, T.M. Hoang, E.M. Bookjans and M.S. Chapman}

\affiliation{School of Physics, Georgia Institute of Technology,
  Atlanta, GA 30332-0430}

\maketitle

{\bf Using squeezed states it is possible to surpass the standard quantum limit of measurement uncertainty by reducing the measurement uncertainty of one property at the expense of another complementary property \cite{Caves81}. Squeezed states were first demonstrated in optical fields \cite{PhysRevLett.55.2409} and later with ensembles of  pseudo spin-1/2 atoms using non-linear atom-light interactions \cite{Polzik10}.  Recently, collisional interactions in ultracold atomic gases have been used to generate a large degree of quadrature spin squeezing in two-component Bose condensates \cite{Nat.464.1165,Nat.464.1170}. For pseudo spin-1/2 systems, the complementary properties are the different components of the total spin vector $\langle \textbf{S} \rangle$, which fully characterize the state on an SU(2) Bloch sphere.
Here, we measure squeezing in a spin-1 Bose condensate, an SU(3) system, which requires measurement of the rank-2 nematic or quadrupole tensor $\langle Q_{ij} \rangle$  as well to fully characterize the state.  Following a quench through a nematic to ferromagnetic quantum phase transition,  squeezing is observed in the variance of the quadratures up to $\mathbf{-8.3_{-0.7}^{+0.6}}$ dB ($\mathbf{-10.3_{-0.9}^{+0.7}}$ dB corrected for detection noise) below the standard quantum limit. This spin-nematic squeezing is observed for negligible occupation of the squeezed modes and is analogous to optical two-mode vacuum squeezing. This work has potential applications to continuous variable quantum information and quantum-enhanced magnetometry.}

The study of many-body quantum entangled states including atomic spin squeezed states is an active research frontier. In addition to being intrinsically fascinating, such states have applications in precision measurements \cite{PhysRevA.46.R6797}, quantum information and fundamental tests of quantum mechanics \cite{RevModPhys.77.513}. Atomic squeezed states were first considered for ensembles of two-level (pseudo spin-1/2) atoms.  For spin-1/2 particles, coherent states of the system are uniquely specified by the components of the total spin vector $\langle \textbf{S} \rangle$, typically illustrated on a SU(2) Bloch sphere. For particles with higher spin, additional degrees of freedom beyond the spin vector are required to fully specify the state.  For spin-1 particles, a natural basis to describe the wavefunction is the SU(3) Cartesian dipole-quadrupole basis, consisting of the three components of the spin vector, $\hat{S}_i$, and the moments of the rank-2 quadrupole or nematic tensor, $\hat{Q}_{ij}$ ($\{i,j\} \in \{x,y,z\}$). In matrix form, the nematic moments can be written $Q_{ij} = S_i S_j + S_j S_i -\frac{4}{3} \delta_{ij}$ \cite{NJP.12.085011}.

Spin-1 atomic Bose-Einstein condensates \cite{PhysRevLett.81.742,JPSJ.67.1822,Stenger99,PhysRevLett.92.140403,Schmaljohann04} provide an exceptionally clean experimental platform to investigate the quantum dynamics of many-body spin systems. They feature controllable quantum phase transitions, well-understood underlying microscopic models, and flexible defect-free geometries. Importantly, it is possible to initialize non-equi\-lib\-ri\-um or excited states of the system and to directly measure both the spin vector and the nematic tensor using standard atomic state manipulation tools.
Law, \emph{et al.}, demonstrated that the spinor interaction can be written as total spin angular momentum, $\lambda \hat{S}^2$ where $\hat{S}^2= \hat{S}^2_x +  \hat{S}^2_y +   \hat{S}^2_z$ \cite{PhysRevLett.81.5257}.
It is natural to describe this many-body system in the second quantized formalism in terms of the mode operators of three Zeeman states (e.g.
$\hat{S}_x = \frac{1}{\sqrt{2}} ( \hat{a}_{1}^{\dag} \hat{a}_{0}
	+\hat{a}_{0}^{\dag} \hat{a}_{-1} + \hat{a}_{0}^{\dag} \hat{a}_{1}
	+ \hat{a}_{-1}^{\dag} \hat{a}_{0} )$).
Here $a_i$ are particle annihilation operators, and we use the single mode approximation where the modes share the same same spatial wavefunction $\phi (\mathbf{r})$ determined by the spin-independent part of the Hamiltonian.
The spinor Hamiltonian describing the collisionally-induced spin dynamics of the condensate and the effects of an applied magnetic field $B$  (by convention along the $z$-axis) can be written (Supplementary Information):
\begin{equation}
	H = \lambda \hat{S}^2 +  \frac{1}{2}q \hat{Q}_{zz}
\label{Hamiltonian}
\end{equation}

\noindent Here
$\lambda$ and $q \propto B^2 $ characterize the inter-spin and Zeeman energies, respectively. At high fields, the system favors nematic ordering of the spins, an eigenstate of $\hat{Q}_{zz}$.  This is a state with $\langle \textbf{S} \rangle=0$ with broken rotational symmetry given by the anisotropy of spin fluctuations, e.g.  $\langle S^2_x \rangle = \langle S^2_y \rangle \neq \langle S^2_z \rangle$ and whose alignment is specified by a time-reversal invariant `director', in this case the $z$-axis. In the Fock basis, $|N_1,N_0,N_{-1}\rangle$, this is just the state with all $N$  atoms in the $m_f = 0 $ state: $|0,N,0\rangle$.
At low fields, the sign of $\lambda$ determines whether the interactions favor a ferromagnetic ($\lambda < 0$, i.e. $^{87}$Rb used in our work) or anti-ferromagnetic ($\lambda > 0$, i.e. $^{23}$Na) ground state with the ground states being maximum and minimum total spin respectively. At intermediate fields, the system undergoes a quantum phase transition between  orders  with a quantum critical point at $q=2|c|$ for the ferromagnetic case, where $c=2N\lambda$ \cite{Nat.443.312}.

\begin{figure*}[t!!!]
\begin{center}
\begin{tabular}{ccccc}
  \begin{tabular}{c}
  	\includegraphics{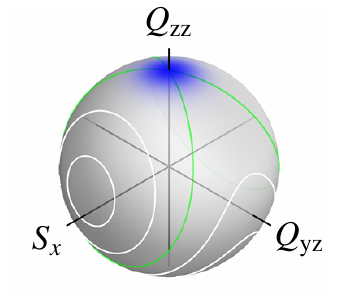} \\
  	\includegraphics{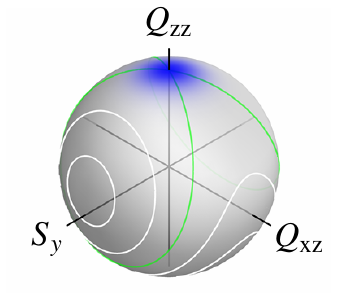} \\
  	\includegraphics{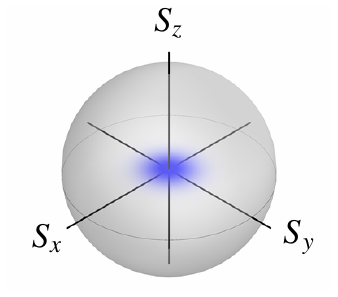}
  \end{tabular} &
  \begin{tabular}{c}
  	\includegraphics{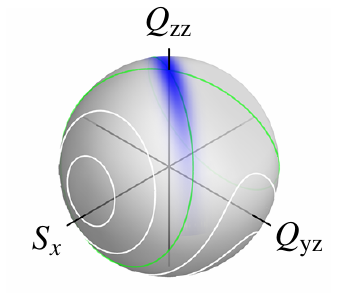} \\
  	\includegraphics{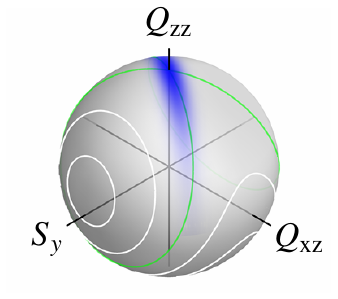} \\
  	\includegraphics{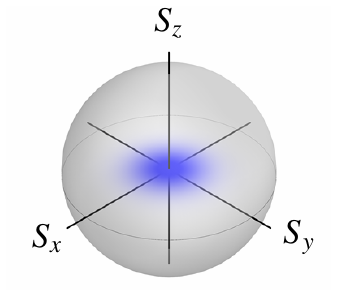}
  \end{tabular} &
  \begin{tabular}{c}
  	\includegraphics{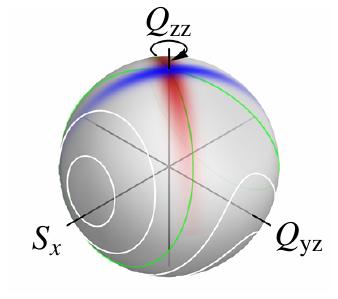} \\
  	\includegraphics{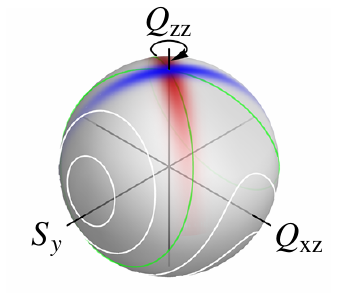} \\
  	\includegraphics{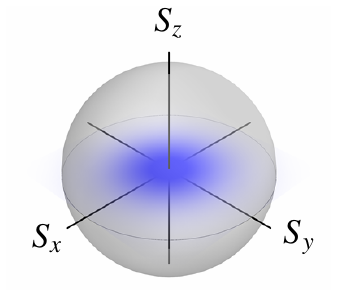}
  \end{tabular} &
  \begin{tabular}{c}
  	\includegraphics{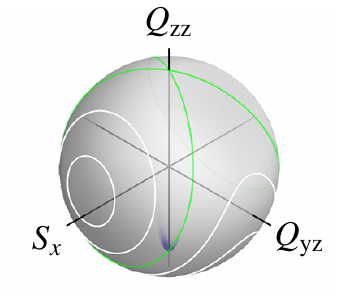} \\
  	\includegraphics{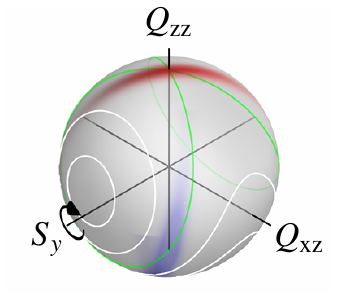} \\
  	\includegraphics{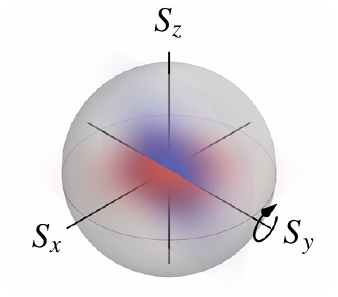}
  \end{tabular} &
  \begin{tabular}{c}
  \includegraphics{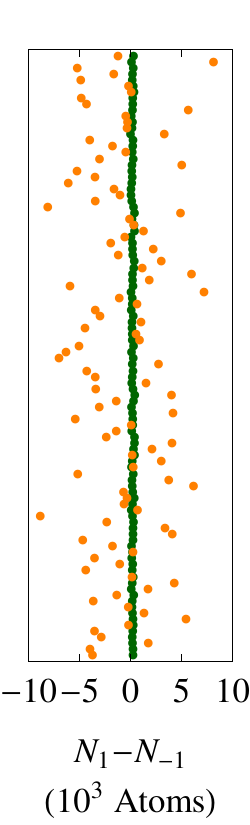}
  \end{tabular} \\
  \textbf{a} & \textbf{b} & \textbf{c} & \textbf{d} & \textbf{e}
  \end{tabular}
\caption{\footnotesize{Illustration of the experimental sequence using semi-classical simulation and quasi-probability distributions.   \textbf{a.} The initial state is a condensate with the atoms prepared in the $m_f = 0$ state. A $N = 30$ atom distribution is used to emphasize features.  \textbf{b.} After $25~\mathrm{ms}$ of evolution, spin-nematic squeezing develops along the separatrix  (green line) in the two upper spheres.  \textbf{c.} A microwave pulse rotates the quadrature phase.  For comparison the state from the previous plot is shown in red in the upper two spheres. \textbf{d.} A $\pi/2$ RF pulse rotates the transverse magnetization $S_x$ into $S_z$.  For comparison the state from the previous plot is shown in red in the lower two spheres.  \textbf{e.}  After the trap is turned off, a Stern-Gerlach field is applied during the TOF expansion and the clouds of atoms are counted using fluorescence imaging.  Shown are points for 100 runs of a squeezed quadrature (green) and an unsqueezed quadrature (orange) from the experimental measurements.}}
\label{SqueezingCartoon}
\end{center}
\end{figure*}

\begin{figure*}[t!!!]
\begin{center}
\begin{tabular}{lll}
	\textbf{a} & \textbf{b} & \textbf{c} \\
	\begin{minipage}{3.45in}
	\includegraphics{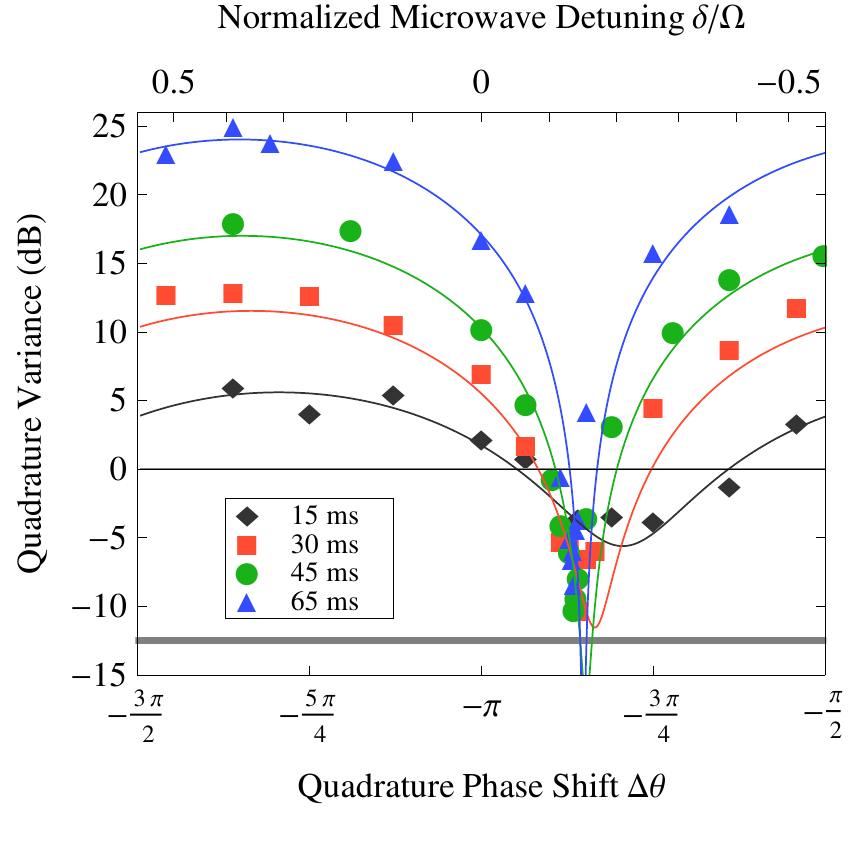}
	\end{minipage}
	&
	\begin{minipage}{1.79in}
		\includegraphics{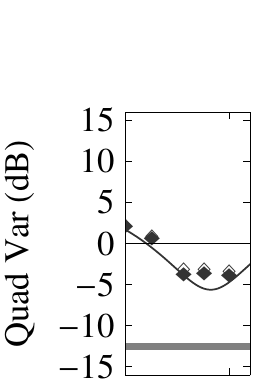}
		\includegraphics{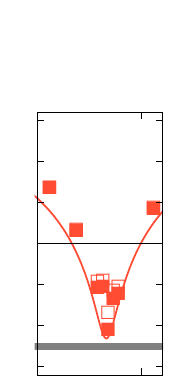} \\
		\includegraphics{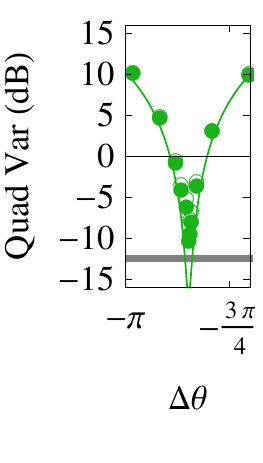}
		\includegraphics{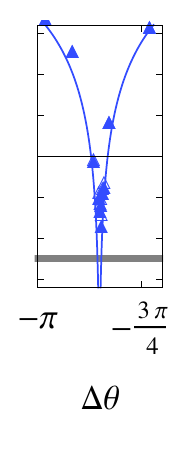}
	\end{minipage}
	&
	\begin{minipage}{1.65in}
	\includegraphics{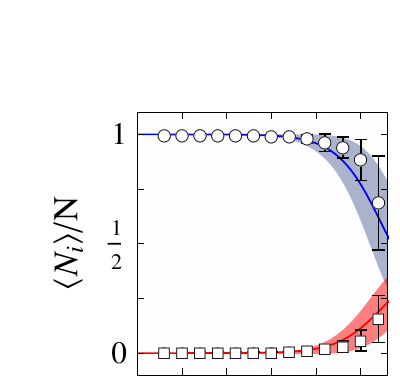} \\
	\includegraphics{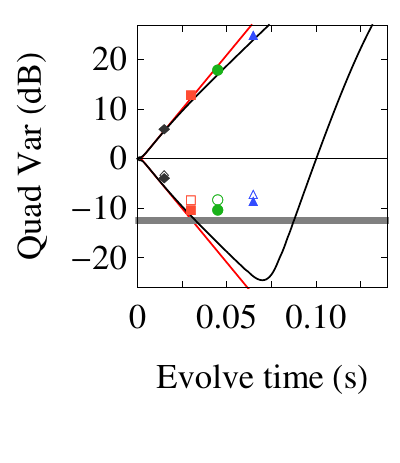}
	\end{minipage}
\end{tabular}
\caption{\footnotesize{Comparison of measured quadrature fluctuations with a fully quantum calculation. \textbf{a.}  Measurement of the quadrature variances for different evolution times and quadrature phase shifts.  The  phase shift is calculated from the microwave detuning normalized to the on-resonance Rabi rate (Supplementary Information).
Estimated errors are approximately the size of the marker for both phase and variance.  Open markers are statistics of the raw data, filled markers have been corrected for PSN.  \textbf{b.}  Detailed view of the maximum squeezing for different evolution times. The phase of data is shifted to match the simulation.   \textbf{c.} Time evolution of the populations and the maximum and minimum quadrature variances.  The squeezing measurment is performed prior to significant population evolution.}}
\label{FockSimVSExp}
\end{center}
\end{figure*}

In spin-1/2 systems, non-linear Hamiltonians such as $S^2_z$ produce squeezing of the spin variables satisfying Heisenberg uncertainty relations, e.g. $\Delta S_x \Delta S_y \geq \frac{1}{2} |\langle  S_z  \rangle|$. While criteria for squeezing and entanglement are now well-established for spin-1/2 particles within a SU(2) framework, there was considerable early discussion about different squeezing definitions  \cite{PhysRevA.47.5138,PhysRevA.50.67}. There has been much less work for higher spin particles with correspondingly higher symmetries.  Squeezing in spin-1 systems has been studied from the perspective of two-mode squeezing \cite{PhysRevLett.85.3987,PhysRevLett.85.3991,PhysRevA.65.033619,PhysRevA.77.023616}, in terms of the Gell-Mann (quark) framework of the SU(3) algebra \cite{You02}, and in terms of spin-nematic measurables \cite{NJP.12.085011}.  Although in each study appropriate phase spaces for squeezing were identified, a comprehensive picture remains to be developed.

The source of squeezing in a spin-1 condensate is the non-linear collisional spin interaction $\mathcal{H}_s = \lambda \hat{\textbf{S}}$$^2$, which reduces to $\hat{\textbf{S}}$$^2$$ \rightarrow \hat{S}^2_x +  \hat{S}^2_y$ for states with $\langle \hat{S}_z \rangle = 0$ of interest here.
This Hamiltonian contains four-wave mixing terms, $\mathcal{H}_{\mathrm{FWM}} = 2 \lambda ( \hat{a}_{0}^{\dag\,2} \hat{a}_{1} \hat{a}_{-1} + \hat{a}_{1}^\dag \hat{a}_{-1}^\dag \hat{a}_{0}^2 )$, which generate squeezing that can be described using a two-mode formalism where the $\hat{a}_{\pm1}$ modes are the signal and idler (Supplementary Information).  Here we prefer an analysis using the commutators of the SU(3) group describing the spin-1 system as it provides more insight. Experimental investigations of spin-1 condensates have been predominantly in the mean-field limit, however first explorations beyond the mean field have been reported  \cite{Lett09b, Leslie09, Klempt10,PhysRevLett.107.210406}.

From the generalized uncertainty relation $ \Delta A  \Delta B \geq \frac{1}{2} | \langle [ \hat{A},\hat{B} ] \rangle |$, only operator pairs with non-zero expectation values for their commutation relations can exhibit squeezing.
For condensates with the atoms in the $m_f=0$ state, only two of the SU(3) commutators have non-zero expectation values
: $\langle 0,N,0| [\hat{S}_x,\hat{Q}_{yz}] |0,N,0\rangle = - 2 i N$ and $\langle 0,N,0| [\hat{S}_y,\hat{Q}_{xz}] |0,N,0\rangle = 2 i N$, leading to the relevant uncertainty relations  $\Delta S_x \Delta Q_{yz} \geq N$ and $\Delta S_y \Delta Q_{xz} \geq N$. For each of these, the uncertainty relationship is between a spin operator and a quadrupole nematic operator; these operators and their commutators define two subspaces.
From these relations, two squeezing parameters are defined in terms of quadratures of the operators:
\begin{equation}
	\xi_{x(y)}^2 = \langle ( \Delta( \cos \theta S_{x(y)}
	+ \sin \theta Q_{yz(xz)} ) )^2 \rangle /N.
\end{equation}
with $\theta$ as the quadrature angle \cite{PhysRevA.65.033619,NJP.12.085011}.  Squeezing within a given SU(2) subspace is indicated by the variance of the quadrature  operator being less than the standard quantum limit (SQL) of $N$ for some value of $\theta$.

The experimental sequence is illustrated in Fig.~1 with the help of the spin SU(2) subspace $\{S_x,S_y,S_z\}$ and both of the subspaces that exhibit squeezing, $\{S_x,Q_{yz},Q_{zz}\}$ and $\{S_y,Q_{xz},Q_{zz}\}$ (Supplementary Information). The initial state of the condensate is shown in Fig.~1a.  It has no spin moment, $\langle \textbf{S} \rangle =0$, but non-zero diagonal quadrupole elements $\langle Q_{ii} \rangle \neq 0$, and uncorrelated equal uncertainties in $S_x, S_y, Q_{xz}, Q_{yz}$.  Subsequent evolution is governed by Eq.~1 in the regime $q \leq 2|c|$. Out of equilibrium dynamics of spin-1 condensates generally exhibit oscillatory behavior in the spin components \cite{NatPhys.1.111,PhysRevA.72.063619}, except near a separatrix (green contour) in phase space where the period diverges \cite{PhysRevA.72.013602}. For our case, the initial state lies on the separatrix, and hence, evolution is wholly dictated by quantum fluctuations corresponding to the initial uncertainties.  These uncertainties evolve exponentially, generating anti-squeezing for a quadrature of $\{S_x,Q_{yz}\}$ aligned along a branch of the separatrix and squeezing for the orthogonal quadrature, as shown in Fig~1b.  Measuring the squeezing requires state tomography involving two SU(3) rotations.  The first is a rotation about $Q_{zz}$ that rotates the quadrature squeezing to align to $S_x$.  Conceptually, $Q_{zz}$ is the generator of the rotation in the squeezed subspaces and the rotation of the quadrature angle is given by the operator $\exp ( i\,\Delta\theta\,Q_{zz} )$. The second is a $\pi /2$ rotation about the $S_y$ axis (in the lab frame) that rotates the fluctuations in $S_x$ into the measurement basis, $S_z = N_1 - N_{-1}$ (Fig.~1d).   These manipulations are equivalent to the homodyne technique used in quantum optics to measure two-mode squeezing where the $\hat{a}_0$ mode is the local oscillator (Supplemental Information).

Identical squeezing also occurs in the degenerate $\{S_y,Q_{xz},Q_{zz}\}$ subspace, which leads to an important, but subtle point. In the lab frame, the system does not confine itself to either of the SU(2) subspaces but rather undergoes rapid Larmor precession  described by rotations about $S_z$ in two other SU(2) subspaces $\{S_x,S_y,S_z\}$ and $\{Q_{yz},Q_{xz},S_z\}$ (not shown).  However, because the squeezing is identical in both subspaces, and the Larmor precession of the spin vector and quadrapole are synchronized, it is not necessary to track the precession in order to measure squeezing.  This has important experimental consequences in that it is not necessary to maintain synchronization with the Larmor rotation in order to perfom quantum state tomography.

\begin{figure*}[tp]
\begin{center}
\begin{tabular}{ll}
	\textbf{a} & \textbf{b}\\
	\includegraphics{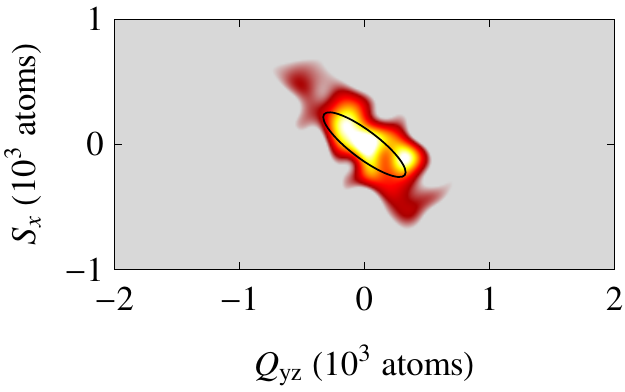} &
	\includegraphics{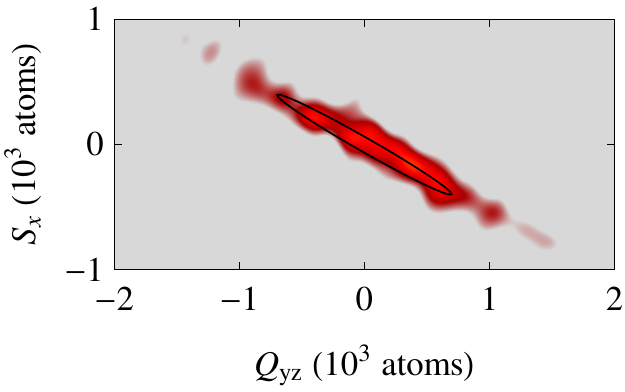} \\
	\textbf{c} & \textbf{d} \\
	\includegraphics{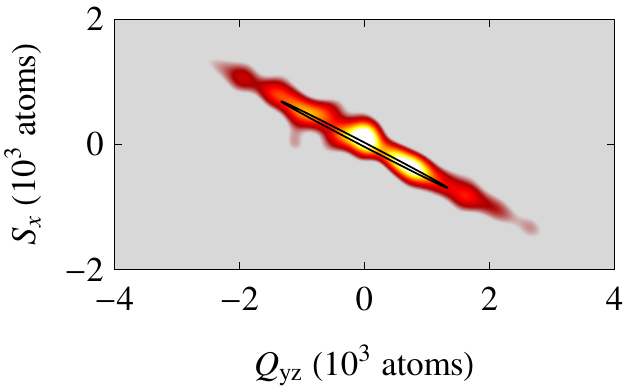} &
	\includegraphics{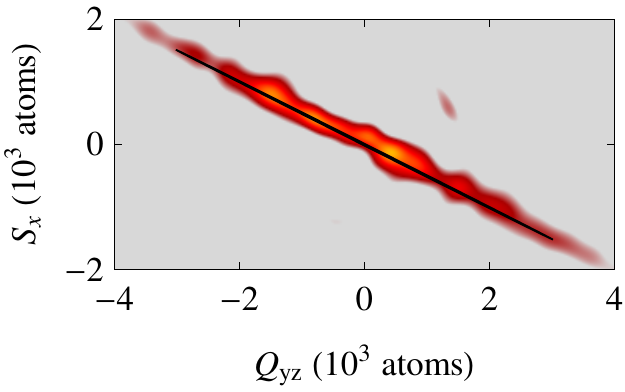}
\end{tabular}
\caption{\footnotesize{Reconstructions of the phase space for different evolution times:  \textbf{a} 15 ms, \textbf{b} 30 ms, \textbf{c} 45 ms, and \textbf{d} 65 ms.  The last two are at half the scale of the first.  The black trace in each is the calculated $1/\sqrt{e}$ uncertainty ellipse from the simulation.}}
\label{Recons}
\end{center}
\end{figure*}

For the experiment, we prepare a  condensate of $N= 45,000$ $^{87}$Rb atoms in the  $|f=1,m_f=0 \rangle$ hyperfine state in a high magnetic field ($2~\mathrm{G}$). The condensate is tightly confined in an optical dipole trap with trap frequencies of $  250~\mathrm{Hz}$.  For these parameters the Thomas-Fermi radii of the condensate are $3.8~\mu\mathrm{m}$ while the spin healing length is $11~\mu\mathrm{m}$ indicating that the condensate is well described by the single mode approximation.  To initiate evolution, the condensate is quenched below the quantum critical point by lowering the magnetic field to a value $210~\mathrm{mG}$ and then allowed to freely evolve for a set time. During this evolution is when the squeezing is developed.
Note during this time there is essentially no population transfer ($<1\%$) from the $m_f=0$ state (Fig.\thinspace\ref{FockSimVSExp}c top), hence this corresponds to squeezed vacuum of the $m_f=\pm1$ modes.
The quadrature rotation is accomplished by shifting the phase of the $m_f=0$ amplitude using a microwave detuned from the clock transition.  This is immediately followed by an RF rotation to rotate $S_x$ into $S_z$.
The trap is then turned off and a Stern-Gerlach field is applied to separate the $m_f$ components during 22~ms time-of-flight expansion. The atoms are probed  for $400~\mu\mathrm{s}$ with three pairs of orthogonal laser beams, and the resulting fluorescence signal is collected by a CCD camera.  Sample $S_z$ measurements are shown in Fig.~1e for two different quadrature angles and show qualitatively both enhanced and suppressed fluctuations.

The experiment cycle is repeated 100 times for each quadrature angle to collect sufficient statistics to determine the fluctuations $\xi_{x}^2$. The measurement results are shown in Fig.\thinspace\ref{FockSimVSExp} for different evolution times. The fluctuations clearly exhibit noise reduced below the SQL ($\xi_{x}^2 <0$ dB) for certain quadrature phases and show increased noise $\pi /2$ radians away.  The measurements are compared to a fully quantum theoretical calculation (Supplemental Information) performed by numerically integrating Eq. (\ref{Hamiltonian}) with a spinor dynamical rate $c/h = -8~\mathrm{Hz}$ ($h$ is Planck's constant) and $q(t)$ determined  experimentally using microwave spectroscopy. The spinor dynamical rate is chosen to match the degree of anti-squeezing observed and is also a good fit to the long time evolution of the populations (Fig.\thinspace\ref{FockSimVSExp}c top) as well as estimates from the trap frequencies and $N$.

As evolution time is increased, the maximum variance shows anti-squeezing that increases in good agreement with the simulation (Fig.\thinspace\ref{FockSimVSExp}c. black lines)  and is approximately exponential with a time constant of $(2|c|/\hbar)^{-1}$ given by the two-mode squeezing model (Fig.\thinspace\ref{FockSimVSExp}c. thin gray lines).  The minimum variance shows squeezing that limits asymptotically after $\sim 30~\mathrm{ms}$ of evolution due to detection noise from a combination of light scattered by the apparatus and the photo-electron shot noise (PSN). The PSN limit is indicated by the gray lines in Fig.\thinspace\ref{FockSimVSExp}.  The maximum observed squeezing is $-8.3_{-0.7}^{+0.6}~\mathrm{dB}$, which is the highest degree of quadrature squeezing observed in any atomic system.  When corrected for the PSN, it is possible to infer a `corrected' squeezing parameter of $-10.3_{-0.9}^{+0.7}~\mathrm{dB}$  that would be obtained with detection improvements.   The phase of maximum squeezing also evolves in time, converging to the phase of the separatrix given by $\cos \theta_s = - q/c-1$ with a small ($\sim 150~\mathrm{mrad}$) discrepancy between the  measured phase of maximum squeezing and the theoretical prediction.
After this initial time of approximately exponential squeezing, the minimum squeezing parameter reaches a turning point due to pump depletion.

 We also reconstruct the phase space distribution of the squeezing for each time (Fig.\thinspace\ref{Recons}) using an inverse Radon transform. (Supplementary Information).  The theoretical squeezing ellipse is shown for comparison.  Just as the theoretical prediction shows the quadrature evolves both in aspect ratio and phase rotation, 
however,  quantitative agreement is limited by the finite number of data sets.  Furthermore the minimum width of parts c and d is limited by the PSN.

In conclusion, we have observed spin-nematic squeezed states created from the free-evo\-lu\-tion of a spin-1 condensate after being quenched through a nematic-fer\-ro\-mag\-net\-ic quantum phase transition.  This work demonstrates quantum dynamics of a mesoscopic SU(3) spin-nematic many-body system and paves the way for further investigations beyond the mean-field limit.   It shows that simple evolution of a spin-1 condensate is a robust way to generate spin-nematic squeezed states with large amounts of squeezing. These states, together with the tools we have demonstrated for their characterization, can be used for quantum metrology of magnetic fields and for atomic clocks.  These experiments demonstrate new paths to explore the fascinating intersections of correlations, entanglement and quantum phase transitions in an exotic quantum magnetic system.

Very recently, two papers related to ours have appeared \cite{Sci.334.773,Nat.480.219}.  These experiments share a similar spin mixing mechanism to generate quantum correlations, though in their case it is in the $f = 2$ hyperfine manifold. For both of these other experiments, the starting point is the observation of strong non-classical fluctuations of $S_z = N_1 - N_{-1}$  due to spin mixing, which we previously reported in Ref. \cite{PhysRevLett.107.210406}.
In Ref. \cite{Nat.480.219}, atomic homodyne detection of the quadrature fluctuations using a technique that is very similar to ours was demonstrated.  In their case, they were not able to show that the fluctuations were smaller than the classical limit, only that the fluctuations were dependent on the local oscillator phase.  Only by subtracting off detection noise do they claim entanglement of the two modes at $4 \pm 17$ atom level.
In Ref. \cite{Sci.334.773}, the number-squeezed $m_f = \pm1$ states are coupled using a sequence of microwave transitions to realize a two-level pseudo spin-$\frac{1}{2}$ system. On the corresponding Bloch sphere, the number squeezed states are completely phase uncorrelated.  By measuring the resulting variance and $4^\mathrm{th}$ moment of $S_z$, they were able to show that the angle could be determined with an uncertainty -1.6 dB below the classical limit.
These systems are in the regime of twin-atom states similar where the $m_f = \pm1$ states are macroscopically populated.  In this regime there is no longer a simple spin-nematic squeezed vacuum interpretation of the squeezing, but it is still possible to measure quantum correlations as these experiments demonstrate.

\bibliography{SqueezeRef2}

\begin{thebibliography}{10}

\bibitem{Caves81}
Caves, C.~M.
\newblock {\em Phys. Rev. D}{ \bf 23}, 1693--1708 (1981).

\bibitem{PhysRevLett.55.2409}
Slusher, R.~E., Hollberg, L.~W., Yurke, B., Mertz, J.~C., and Valley, J.~F.
\newblock {\em Phys. Rev. Lett.}{ \bf 55}, 2409--2412 (1985).

\bibitem{Polzik10}
Hammerer, K., S\o{}rensen, A.~S., and Polzik, E.~S.
\newblock {\em Rev. Mod. Phys.}{ \bf 82}, 1041--1093 (2010).

\bibitem{Nat.464.1165}
Gross, C., Zibold, T., Nicklas, E., Est\`eve, J., and Oberthaler, M.~K.
\newblock {\em Nature}{ \bf 464}, 1165--1169 (2010).

\bibitem{Nat.464.1170}
Riedel, M.~F., B\"ohi, P., Li, Y., H\"ansch, T.~W., Sinatra, A., and Treutlein,
  P.
\newblock {\em Nature}{ \bf 464}, 1170--1173 (2010).

\bibitem{PhysRevA.46.R6797}
Wineland, D.~J., Bollinger, J.~J., Itano, W.~M., Moore, F.~L., and Heinzen,
  D.~J.
\newblock {\em Phys. Rev. A}{ \bf 46}, R6797--R6800 Dec  (1992).

\bibitem{RevModPhys.77.513}
Braunstein, S.~L. and van Loock, P.
\newblock {\em Rev. Mod. Phys.}{ \bf 77}, 513--577 Jun  (2005).

\bibitem{NJP.12.085011}
Sau, J.~D., Leslie, S.~R., Cohen, M.~L., and Stamper-Kurn, D.~M.
\newblock {\em New J. Phys.}{ \bf 12}(8), 085011 (2010).

\bibitem{PhysRevLett.81.742}
Ho, T.-L.
\newblock {\em Phys. Rev. Lett.}{ \bf 81}(4), 742--745 (1998).

\bibitem{JPSJ.67.1822}
Ohmi, T. and Machida, K.
\newblock {\em J. Phys. Soc. Jpn}{ \bf 67}(6), 1822--1825 (1998).

\bibitem{Stenger99}
Stenger, J., Inouye, S., Stamper-Kurn, D.~M., Miesner, H.-J., Chikkatur, A.~P.,
  and Ketterle, W.
\newblock {\em Nature}{ \bf 396}, 345 (1999).

\bibitem{PhysRevLett.92.140403}
Chang, M.-S., Hamley, C.~D., Barrett, M.~D., Sauer, J.~A., Fortier, K.~M.,
  Zhang, W., You, L., and Chapman, M.~S.
\newblock {\em Phys. Rev. Lett.}{ \bf 92}(14), 140403 (2004).

\bibitem{Schmaljohann04}
Schmaljohann, H., Erhard, M., Kronj\"ager, J., Kottke, M., van Staa, S.,
  Cacciapuoti, L., Arlt, J.~J., Bongs, K., and Sengstock, K.
\newblock {\em Phys. Rev. Lett.}{ \bf 92}, 040402 (2004).

\bibitem{PhysRevLett.81.5257}
Law, C.~K., Pu, H., and Bigelow, N.~P.
\newblock {\em Phys. Rev. Lett.}{ \bf 81}(24), 5257--5261 (1998).

\bibitem{Nat.443.312}
Sadler, L.~E., Higbie, J.~M., Leslie, S.~R., Vengalattore, M., and
  Stamper-Kurn, D.~M.
\newblock {\em Nature}{ \bf 443}, 312--315 (2006).

\bibitem{PhysRevA.47.5138}
Kitagawa, M. and Ueda, M.
\newblock {\em Phys. Rev. A}{ \bf 47}(6), 5138--5143 (1993).

\bibitem{PhysRevA.50.67}
Wineland, D.~J., Bollinger, J.~J., Itano, W.~M., and Heinzen, D.~J.
\newblock {\em Phys. Rev. A}{ \bf 50}(1), 67--88 (1994).

\bibitem{PhysRevLett.85.3987}
Pu, H. and Meystre, P.
\newblock {\em Phys. Rev. Lett.}{ \bf 85}(19), 3987--3990 Nov  (2000).

\bibitem{PhysRevLett.85.3991}
Duan, L.-M., S\o{}rensen, A., Cirac, J.~I., and Zoller, P.
\newblock {\em Phys. Rev. Lett.}{ \bf 85}(19), 3991--3994 Nov  (2000).

\bibitem{PhysRevA.65.033619}
Duan, L.-M., Cirac, J.~I., and Zoller, P.
\newblock {\em Phys. Rev. A}{ \bf 65}(3), 033619 (2002).

\bibitem{PhysRevA.77.023616}
Mias, G.~I., Cooper, N.~R., and Girvin, S.~M.
\newblock {\em Phys. Rev. A}{ \bf 77}(2), 023616 (2008).

\bibitem{You02}
M\"ustecapl\ifmmode \imath \else \i \fi{}o\ifmmode~\breve{g}\else \u{g}\fi{}lu,
  O.~E., Zhang, M., and You, L.
\newblock {\em Phys. Rev. A}{ \bf 66}, 033611 (2002).

\bibitem{Lett09b}
Liu, Y., Gomez, E., Maxwell, S.~E., Turner, L.~D., Tiesinga, E., and Lett,
  P.~D.
\newblock {\em Phys. Rev. Lett.}{ \bf 102}, 225301 (2009).

\bibitem{Leslie09}
Leslie, S.~R., Guzman, J., Vengalattore, M., Sau, J.~D., Cohen, M.~L., and
  Stamper-Kurn, D.~M.
\newblock {\em Phys. Rev. A}{ \bf 79}(4), 043631 (2009).

\bibitem{Klempt10}
Klempt, C., Topic, O., Gebreyesus, G., Scherer, M., Henninger, T., Hyllus, P.,
  Ertmer, W., Santos, L., and Arlt, J.~J.
\newblock {\em Phys. Rev. Lett.}{ \bf 104}(19), 195303 (2010).

\bibitem{PhysRevLett.107.210406}
Bookjans, E.~M., Hamley, C.~D., and Chapman, M.~S.
\newblock {\em Phys. Rev. Lett.}{ \bf 107}, 210406 Nov  (2011).

\bibitem{NatPhys.1.111}
Chang, M.-S., Qin, Q., Zhang, W., and Chapman, M.~S.
\newblock {\em Nat. Phys.}{ \bf 1}, 111--116 (2005).

\bibitem{PhysRevA.72.063619}
Kronj\"ager, J., Becker, C., Brinkmann, M., Walser, R., Navez, P., Bongs, K.,
  and Sengstock, K.
\newblock {\em Phys. Rev. A}{ \bf 72}, 063619 Dec  (2005).

\bibitem{PhysRevA.72.013602}
Zhang, W., Zhou, D.~L., Chang, M.-S., Chapman, M.~S., and You, L.
\newblock {\em Phys. Rev. A}{ \bf 72}(1), 013602 (2005).

\bibitem{Sci.334.773}
L\"ucke, B., Scherer, M., Kruse, J., Pezzé, L., Deuretzbacher, F., Hyllus, P.,
  Topic, O., Peise, J., Ertmer, W., Arlt, J., Santos, L., Smerzi, A., and
  Klempt, C.
\newblock {\em Science}{ \bf 334}(6057), 773--776 (2011).

\bibitem{Nat.480.219}
Gross, C., Strobel, H., Nicklas, E., Zibold, T., Bar-Gill, N., Kurizki, G., and
  Oberthaler, M.~K.
\newblock {\em Nature}{ \bf 480}, 219--223 (2011).

\end{thebibliography}
\bibliographystyle{nature}

{\bf Acknowledgements}  We would like to thank D.M. Stamper-Kurn for bringing our attention to Ref. \cite{NJP.12.085011} and for suggesting these investigations. We would like to thank T.A.B. Kennedy, C.A.R. S\'a de Melo, and J.L. Wood for discussions and  A. Zangwill for suggestions about the manuscript.

{\bf Author Contributions}  C.D.H. and M.S.C jointly conceived the study. C.D.H., C.S.G and T.M.H performed the experiment and analysed the data. E.M.B. developed the imaging system. C.D.H. developed essential theory and carried out the simulations. M.S.C supervised the work.

\end{document}